\documentclass[journal]{IEEEtran}%

\usepackage[caption=false,font=footnotesize,labelfont=sf,textfont=sf]{subfig}

\usepackage{balance}

\ifCLASSINFOpdf
   \usepackage[pdftex]{graphicx}
\else
   \usepackage[dvips]{graphicx}
\fi
\begin{document}
%
\title{Device Identification in Blockchain-Based \\Internet of Things}
%
%
%

\author{Ali Dorri,~\IEEEmembership{Member,~IEEE,}
        Clemence Roulin, {}Shantanu Pal,~\IEEEmembership{Member,~IEEE,} Sarah Baalbaki, \\
        Raja Jurdak, \IEEEmembership{Senior Member,~IEEE,}
        and~Salil S. Kanhere,~\IEEEmembership{Senior Member,~IEEE}

\thanks{A. Dorri, S. Pal, and R. Jurdak are with the School of Computer Science, Queensland University of Technology, Brisbane, QLD 4000, Australia (e-mail: ali.dorri@qut.edu.au, shantanu.pal@qut.edu.au, r.jurdak@qut.edu.au).}

\thanks{C. Roulin is with CSIRO Data61, Pullenvale QLD 4069 Australia.}

\thanks {S. S. Kanhere is with the School of Computer Science and Engineering, University of New South Wales, Sydney, NSW 2052, Australia (e-mail: salil.kanhere@unsw.edu.au).}

\thanks{S. Baalbaki is with the American University of Beirut, Beirut Campus, Beirut 110236, 
Lebanon (e-mail: sib09@mail.aub.edu).}}
\maketitle

\begin{abstract}
In recent years blockchain technology has received tremendous attention. Blockchain users are known by a changeable Public Key (PK) that introduces a level of anonymity, however, studies have shown that anonymized transactions can be linked to deanonymize the users. Most of the existing studies on user de-anonymization focus on monetary applications, however, blockchain has received extensive attention in non-monetary applications like IoT. In this paper we study the impact of de-anonymization on IoT-based blockchain. We populate a blockchain with data of smart home devices and then apply machine learning algorithms in an attempt to classify transactions to a particular device that in turn risks the privacy of the users. Two types of attack models are defined: (i) informed attacks: where attackers know the type of devices installed in a smart home, and (ii) blind attacks:  where attackers do not have this information. We show that machine learning algorithms can successfully classify the transactions with  90\%  accuracy. To enhance the anonymity of the users, we introduce multiple obfuscation methods which include combining multiple packets into a transaction, merging ledgers of multiple devices, and  delaying transactions. The implementation results show that these obfuscation methods significantly reduce the  attack success rates to  20\% to 30\% and thus enhance user privacy. 
\end{abstract}


%
\IEEEpeerreviewmaketitle

\section{Introduction}
\label{sec:intro}
Blockchain is a disruptive technology that has received tremendous attention from practitioners and academia   due to its salient features including auditability, immutability, decentralization, and anonymity~\cite{dai2019blockchain}. In blockchain, a transaction forms the basic communication primitive that allows nodes to exchange information or read/write in the blockchain. It is possible that there exists dependencies between  transactions where certain fields generated in one transaction (outputs) are refrenced in another transaction as inputs~\cite{perera2020blockchain}. Particular nodes in the network, known as miners, collate multiple transactions and form a block which is appended to the blockchain by following  a consensus algorithm. Using the consensus algorithm, e.g., Proof of Work (PoW) \cite{nakamoto2008bitcoin}, participating nodes build a trusted network over untrusted participants. 

New transactions and blocks are broadcast and verified by all participating nodes that eliminates the need for central authorities and introduces distributed management of trust.  Each block includes  the hash of its previous block in the ledger that  ensures blockchain immutability. The modification of the block content, i.e.,  the transactions, is impossible since    the hash maintained in the subsequent block will not match with the hash of the modified block. The transactions are permanently stored in the public immutable blockchain where any node can retrieve in any point of time that introduces high auditability~\cite{8731639}.

The transactions are cryptographically sealed using public/private keys. The Public Key (PK\textsuperscript{+}) used in each transaction is employed as the identity of the transaction generator. This introduces a level of anonymity for the blockchain users as their real identity remains unknown to the participating nodes.  To enhance their anonymity, the users may change their PK\textsuperscript{+} for each new transaction. The anonymity of the users in crytocurrencies, the first application of blockchains,  has been widely studied in the literature \cite{moser2013anonymity} \cite{fanti2017deanonymization}. The studies suggest that the malicious nodes can deanonymize a user by:  (i) linking multiple transactions with the same PK\textsuperscript{+},   (ii) classifying  transactions with different PK\textsuperscript{+}s based on particular metrics, e.g., the flow of inputs/outputs, and linking them to a user, (iii)  analyzing real-time network traffic , or (iv) employing off-chain information, e.g., the information available in the website of a company. 

In recent years, there has been extensive research on adopting and applying blockchain in  non-monetary applications including  the network of billions of connected devices that form the Internet of Things (IoT) \cite{dorri2019lsb} \cite{kumar2022survey}. Due to its salient features as outlined above, blockchain has the potential to address security, anonymity, and centralization  challenges and enhance the auditability and transparency of the conventional IoT frameworks. In IoT each user owns a number of devices that collect and share  data with Service Providers (SP) and/or other users to offer personalized services to the user.  IoT devices encrypt data in transit to  increase security.  Exposure of the  user's activity patterns results in serious privacy and security concerns, e.g., an attacker may infer the hours that a home is occupied by monitoring the pattern of transactions generated by motion sensors~\cite{pal2020security}.  In a blockchain-based IoT, the transactions reflect the activities of a participating IoT device which potentially may expose the user activity pattern and thus compromise their privacy.  Blockchain transactions may not contain the raw data shared by the IoT devices; however, an attacker can still expose the  user's activity without accessing data,  by monitoring the temporal pattern of stored encrypted communications, i.e., transactions, of IoT devices~\cite{apthorpe2017smart}. An attacker with the intention of revealing a user's activities must first determine the types of IoT devices in the user's premises. The combination of user deanonymisation and IoT device identification can therefore be a powerful tool for an attacker to determine a user's identity and activities. 

Device identification in blockchain is conceptually similar to device classification in IoT \cite{miettinen2017iot} that is a security tool employed by the network managers to identify the type of devices installed in their site and thus detect malicious unauthorized devices and traffic. 
Unlike conventional IoT, where device classification is employed to enhance network security, when applied to blockchain-based IoT, it can lead to privacy violations. Device classification in blockchain is different to when applied to conventional IoT due to the following reasons: 
\begin{itemize}
	\item Blockchain is an immutable system that makes it impossible to remove or modify transactions that are previously stored in the chain.  This potentially turns  blockchain into a permanent database that exposes the history of the communications and activities of the IoT participants. However, in conventional IoT, the historical communications of a devices is only stored by the device and SP. To read communications, one shall perform attacks, e.g., man-in-the-middle attack to access real-time IoT traffic. 
	
	\item Blockchain transactions are broadcast to the network and the transaction recipient (if any) is identified by the  PK\textsuperscript{+}. Thus, the IP address of the transaction generator is not stored in the blockchain. In conventional IoT data classification methods, the IP address is employed as a source to classify the devices, e.g., the destination IP address reveal information about the recipient of the transaction.
	\item Transactions contain PK and record the interactions between the participants, while no network specific field or data is stored in blockchain. However, in conventional IoT device classification methods real-time access to the network is needed. Thus, network layer packets can also be accessed which can facilitate device classification.
\end{itemize}

The \textit{major contribution} of this paper is to study the possibility of device identification in blockchain-based IoT.  To the best of our knowledge, this is the first attempt to \textit{identify types of devices} in an IoT blockchain context. We use a smart home setting as a representative use case to study device classification in IoT. We utilize the smart home traffic dataset available in \cite{UNSWIoTDataset} to populate a blockchain. Each transaction in the blockchain corresponds to a particular communication in the dataset with the same timestamp. We study the success rate of classifying transactions corresponding to a particular device and identifying the device type based on the transaction temporal pattern.  The attacker's aim is to match temporal signatures observed in the transaction ledger to known IoT device temporal patterns in order to identify the types of involved devices.  As the attack is a pattern matching problem, we use machine learning, particularly decision trees, to model the attacker. We define two attack models: (i) \textit{informed attack} where the attacker knows the number and type of  devices that are in the smart home, but does not know the mapping of devices to identities in the ledger. This attack is highly unlikely as it requires an attacker to monitor the user's device acquisitions over a long period of time, yet we include it as a worst case scenario for studying user activity privacy risk, and (ii) \textit{blind attack} where the attacker has no specific information about the number and type of devices in the smart home. Our results show that up to 90\% classification accuracy of the type and number of devices in the ledger can be achieved. 

To protect against this attack, we propose four timestamp obfuscation methods: combining multiple packets into a transaction, merging ledgers of multiple devices,  randomly delaying transactions, and adding fake transactions to alter the temporal pattern of transactions. We study the impact of the proposed methods on attack success.  The timestamp obfuscation methods can reduce attack success rates to between 20-30\%.

The rest of the paper is organized as follows. In Section~\ref{sec:lit-review}, we present the related work. In Section~\ref{sec:deviceClassification}, we provide a detailed discussion on device classification in blockchain-based IoT.  System evaluation and results are illustrated in Section~\ref{sec:evaluation}. Finally, Section~\ref{sec:conclusion} concludes the paper and outlines  future work.

\section{Related Work}\label{sec:lit-review}
In this section, we present a literature review on the user deanonymization in blockchain-based systems. As outlined in Section \ref{sec:intro}, blockchain was first proposed in cryptocurrencies. The existing literature in user deanonymization in blockchain mainly focuses on cryptocurrencies. Thus, we discuss such methods in this section.

\subsection{User Anonymity in Blockchain}
To study the anonymity of the users, it is critical to consider the following properties \cite{amarasinghe2019survey}:
\begin{itemize}
	\item \textit{Anonymity of an entity}  is the ability to identify a particular node within a group of nodes, that are known as anonymity set. 
	\item \textit{Recipient  anonymity} is the ability to identify a particular node as the recipient of a (group of) transaction(s).
	\item \textit{Unlinkability} refers to the impossibility of linking multiple transactions to a particular user.   
	\item \textit{Untraceability} refers to the impossibility of tracking a transaction to the transaction generator. 
	\item \textit{Hidden transaction value} refers to the impossibility of finding the actual input/output value of a transaction.    
\end{itemize}
The anonymity of a blockchain-based system depends on the degree to which it satisfies the outlined properties~\cite{dorri2019lsb-new}. In the rest of this subsection we study the existing methods in the literature to deanonymize blockchain users.  

\textbf{Active Interaction:}  In this method, the malicious nodes directly interact with the target node using transactions and capture the real-time  information about the target. As an example, a malicious node may pay the price of a good using cryptocurrency and gather information about the seller, e.g., their IP address, PK, etc. The authors in \cite{meiklejohn2013fistful}  attempt to deanonymize Bitcoin users by directly communicating with 31 service providers  (SPs) that include mining pools, wallet software providers, bank exchanges, vendors and gambling sites using 344 transaction leading to 16,086,073 transactions in total in the network. Upon receipt of a transaction from the SPs, the PK\textsuperscript{+} of the transaction generator is recorded. The sender sends multiple transactions to the same SP to increase the number of collected PK\textsuperscript{+}. The collected information is then fed into a transaction analysis algorithm that employs the following two heuristics to analyze the transactions: 
\begin{itemize}
	\item The inputs of a transaction belong to the same user. 
	\item In case a transaction has only one input and one output, both input and output belong to the same user. Such a transaction is  known as a one-time change address. 
\end{itemize}
Utilizing the first heuristic the network is clustered into 5,579,176 clusters of users. The second heuristic is then applied in the clusters leading to 3,384,179 distinct clusters where the identity of 2,197 users was revealed. Particular transactions are linked to a particular SP, e.g., 20  clusters belong to Mt Gox mining pool. 

\textbf{Analyzing Network Traffic:} In this method, the malicious node sniffs the packets exchanged in the underlying peer-to-peer network to capture IP address associated with a transaction.  This in turn enables the malicious nodes to identify multiple transactions generated by the same node and assists  in user deanonymization. The blockchain participants can employ Tor \cite{tor} to hide their IP address. However, the authors in~\cite{biryukov2014deanonymisation} demonstrate that malicious nodes can classify transactions corresponding to a  user based on their IP address even if they employ Tor. Malicious nodes can also identify the users that are connected through the same ISP and thus have the same IP address. Malicious nodes connect to the entry points, i.e., the nodes that connect new nodes to the blockchain, and eavesdrop  the IP address of the newly joined nodes. To increase the attack's success rate, the malicious nodes connect to as many entry points as possible. The malicious node was successful in identifying 11\% of the transactions in Bitcoin.  

\textbf{Analyzing Transactions: } In this method, the malicious node classifies transactions to identify the transactions corresponding to the same user. To classify transactions, malicious nodes construct the following graphs \cite{conti2018survey}: 
\begin{itemize}
	\item Transaction graph: blockchain can be represented as an acyclic graph \textit{G={T,E}}, where \textit{T} is a set of transactions and \textit{E} represents a set edges. The edges represent the flow of inputs/outputs between transactions. 
	\item Address graph: the flow of inputs/outputs in a transaction graph can be represented as an address graph \textit{AG={P,E`}} where \textit{P} is the set of blockchain addresses, i.e., PK\textsuperscript{+}, and \textit{E`} is the set of edges that represents the relationship between addresses.
	\item Entity graph: the PK\textsuperscript{+}s corresponding to a user can be classified by constructing the entity graph \textit{EG={U,E``}} where \textit{U} is a set of PK\textsuperscript{+}s belonging to the  user \textit{U} and \textit{E``} is the edge connecting the users. Malicious nodes use the address graph along with some heuristics, examples discussed earlier in this section, to classify user transactions.
\end{itemize}

In \cite{androulaki2013evaluating} the authors classify the blockchain users by forming the entity graph. First, the address graph is constructed as outlined above.  To evaluate the proposed method, the authors mimic the shopping behavior of staff and students in a university. The profile of  40\% of the participants was constructed with 80\% accuracy.  

\textbf{Using Off-Chain Information:} In this method, the malicious nodes attempt to gather information about a particular user, i.e., PK\textsuperscript{+}, from off-chain information  sources. A user may reveal their PK\textsuperscript{+} in a forum discussion or companies may reveal their  PK\textsuperscript{+}s on their website. The authors in \cite{goldfeder2018cookie} employ cookies set by third parties in online shopping websites to track the PK\textsuperscript{+} used by a users. The results demonstrate that the malicious entity was able to identify 20 out of 25 users.

\subsection{Enhancing User Anonymity}\label{sub:sec:enhancingAnonymity}
In this section, we study the existing methods to improve the user anonymity in blockchain. In this paper, we categorize the existing solutions in two groups which are: (i)  mixing services and (ii) cryptographic methods. \par

Mixing services are widely employed by various cryptocurrencies to improve the anonymity of the users   \cite{feng2019survey}. The nodes that wish to improve their privacy pay a specific value, say \textit{x}, to the mixing service. We represent the  input values to the mixing service as \textit{PK\textsubscript{1}, PK\textsubscript{2},...,PK\textsubscript{n}}. The mixing service then shuffles the input coins and randomly assigns them to new set of keys \textit{PK\textsuperscript{'}\textsubscript{1}, PK\textsuperscript{'}\textsubscript{2},...,PK\textsuperscript{'}\textsubscript{n}} which is then distributed among the nodes, that have initially contributed. Each node shall receive the same amount of coin they paid to the mixing service. However, this may facilitate linking the new keys to the previous key. Say only one of the input values to the mixing service was 10 coins, and thus the only output with 10 coins will belong to  the same node. To address this challenge, in the mixing services  all input transactions as well as output ones have the same value.  \par

Cryptographic methods generally employ cryptographic concepts to  enhance the anonymity of the users \cite{amarasinghe2019survey}. The authors in  \cite{miers2013zerocoin} introduce ZeroCoin that employs zero knowledge proof for spending coins while breaking the historical link between the transactions owned by the same user. In this method, all transactions have to use a fixed coin value.\par

To the best of our knowledge, the existing solutions to study and improve the user anonymity focus on cryptocurrencies. The wide spread adoption of blockchain in IoT applications motivates the need to study anonymity in IoT applications. The key difference in such applications is that transactions typically have recurring temporal signatures that are specific to device types and manufacturers, which can help an attacker infer the device types. We next study the impact of user deanonymization in IoT-based blockchain.

\section{Device Classification in Blockchain-based IoT}\label{sec:deviceClassification}
In this section, we outline the details of device classification in blockchain-based IoT. To mimic the actual transaction pattern generated by IoT devices, we populate a blockchain using  a real-world smart home traffic dataset as discussed in Section \ref{sub:sec:blockchain-setup}. Next, we discuss  the attack models in Section \ref{sub:sec:attack-model}. To enhance the anonymity of the users, we propose multiple timestamp obfuscation methods, which are outlined in Section \ref{sub-sec:time-obfuscation}. 

\subsection{Modeling an IoT-based Blockchain}\label{sub:sec:blockchain-setup}
\begin{figure}
	\centerline
	{\includegraphics[scale=.45]{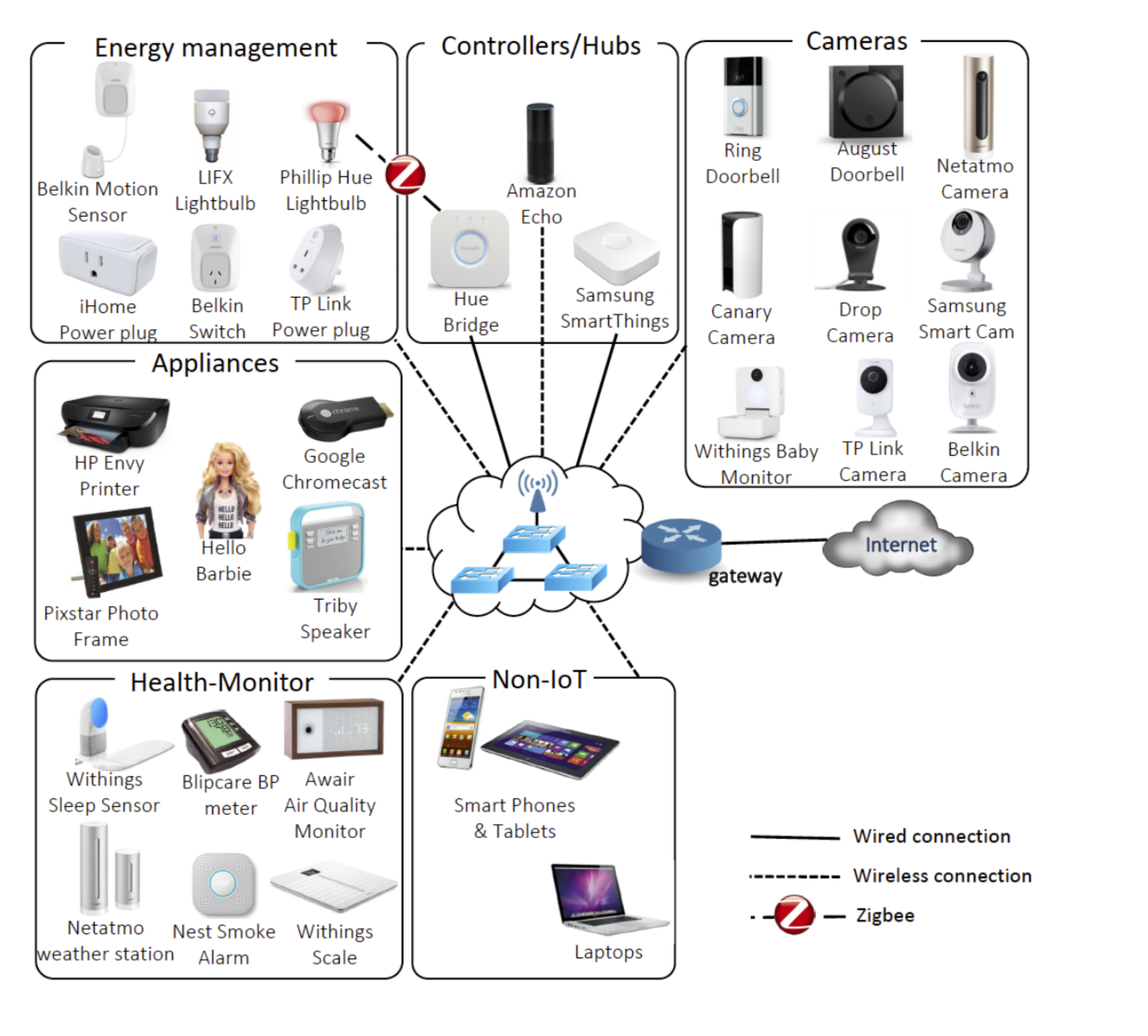}}
	\caption{The overall structure of the smart home dataset \cite{UNSWIoTDataset}.}
	\label{fig:overal-smart home}
\end{figure}

In this section, we discuss the process of populating the blockchain database. We use a smart home application as a representative case on an IoT applications. We populate a blockchain based on the smart home network traffic dataset available at \cite{UNSWIoTDataset}. The dataset contains the traffic of a real smart home network for a period of two weeks that consists of 30 IoT devices. The type of the devices installed in the smart home are shown in  Figure   \ref{fig:overal-smart home}. The devices use a PK\textsuperscript{+} as their identity  which is changed per transaction to enhance the anonymity level.\par 

For each communication in the smart home dataset, a corresponding transaction is  stored in the blockchain. Each transaction is structured as $< T\textsubscript{ID}, Prv\textsubscript{T\textsubscript{ID}}, timestamp, Output,  pk,  sign>$ where \textit{T\textsubscript{ID} } is the unique identifier of the transaction that essentially is the hash of the transaction content. The identity of the previous transaction generated by the same device is stored in \textit{Prv\textsubscript{T\textsubscript{ID}}} which potentially chains transactions generated by the same device.  \textit{Prv\textsubscript{T\textsubscript{ID}}}  also protects against \textit{sybil attack} where a malicious node pretends to be multiple nodes by generating fake transactions  \cite{dorri2019lsb}.   The time when a packet is generated by a device is stored in \textit{timestamp} field. The hash of the \textit{PK\textsuperscript{+}} that will be used in the next transaction is stored in \textit{output} field. The last two fields are  PK\textsuperscript{+} of the transaction generator and its corresponding signature. In case the communication, in the dataset, corresponding to the transaction involves data, the transaction generator  signs the hash of the data and populates in \textit{sign} field, otherwise, the hash of the transaction is signed. Storing data in the blockchain incurs the  packet and memory overhead and reduces blockchain scalability, thus most existing  blockchains store  data in an off-chain database while the hash of the data is stored in blockchain to maintain data integrity. \par

The smart home dataset also consists of network management packets, e.g., SMTP. Note that in conventional blockchain settings, such information is not stored in the blockchain and  only the interactions between the devices is stored. Thus, we disregard network management packets in the dataset.  We also intentionally disregard the blockchain consensus process as the main objective of the attacker is to classify smart home devices using blockchain database that is not impacted by the underlying consensus algorithm. A single node functions as validator and stores transactions in the blockchain in the form of blocks.  We assume that  devices change their Public Key  \textit{PK\textsuperscript{+}} per each transaction they generate to enhance their anonymity.  \par 

The attacker applies machine learning algorithms in the blockchain database to classify devices as outlined in the next section. 

\subsection{Attack Models}\label{sub:sec:attack-model}
As shown in  Table \ref{tab:devices}, the pattern of transactions mostly represent a sequence of in-order numbers. Different patterns share some features, e.g., a separation of 0.207s is found for both the Smarts Things and the Nest smoke alarm. Such patterns represent a classification problem that can be represented by decision trees. Thus, the machine learning algorithms, employed by the attacker, use decision trees to analyze the pattern of transactions in the blockchain and classify devices. The attacker can read the stored transactions and blocks in the blockchain, but cannot decrypt the data associated with the transactions without the corresponding private key.  The attacker trains the machine learning algorithm in a testnet that is a local network consisting the smart home devices. Depending on the number of devices in the testnet,  the ability of the attacker to detect the devices varies. We study the following two attack models:\par 

\begin{table}
\centering
\caption{Inter-packet temporal patterns for devices.}
\scalebox{.97}{
\begin{tabular}{|l|l|} \hline
	\textbf{Device} & \textbf{Patterns of Frequent Time Separation (in sec)} \\ \hline [0.5ex]
	Smart\_Things & 0.207 then 58 then 0.207 then 58 …\\ \hline
	Amazon\_Echo & 0.217 then 30 then 0.004 then 30 …\\ \hline
	TPLink\_Camera & 0.12 then 61 then 0.12 then 61 …\\ \hline
	Samsung\_Camera & 0.165 then 30 then 0.165 then 30 …\\ \hline
	Drop\_Camera & 1.03 or 0.2\\ \hline
	Insteon\_Camera2	& 9x$<$0.0001 then 0.216 then 300 ...\\ \hline
	Baby\_Monitor & 600 then 0.28 then 600 then 0.28 …\\ \hline
	TPLink\_Smartplug & 0.24 then 236 then 0.24 then 236 …\\ \hline
	TPLink\_Smartplug & 0.12 then 236 then 0.12 then 236 …\\ \hline
	iHome & 60 then 0.205 then 60 then 0.205\\ \hline
	Nest\_Smockalarm	& 0.207 then 0.015 then 0.207 then 0.015 …\\ \hline
	Netatmo\_Weather	& 1.72 then 0.33 then 1.72 the 0.33 …\\ \hline
	Sleep\_Sensor & 10 then 0.276 then 10 then 0.276 …\\ \hline
	Lifx\_Smartbulb & 1.92 or 60\\ \hline
	Triby\_Speaker & 120 – 0.3 - 120 – 0.3 - 56 – 0.3 …\\ \hline
	Pix\_Photoframe & 0.31 or $>$=0.3 then 65 then 650\\ \hline
	HP\_Printer & 90\\ \hline
\end{tabular}
}
\label{tab:devices}

\end{table}

\begin{itemize}
	\item \textbf{Informed Attack}: In this attack, the attacker knows the type, including the manufacturer, of devices installed in a smart home.  The entire range of devices in the blockchain must be represented in the  trained  algorithm, thus, we employ  10-fold cross validation analysis to model the informed attacker.  As  the attacker  knows the type of devices in the smart home, he can  acquire similar devices and collect sufficient  network traffic data to populate a comprehensive training set. While this attack is unlikely in practical scenarios, as it  requires the attackers to monitor the user's sensor acquisitions of a long period of time prior to launching the attack, we include it as a worst case scenario for activity privacy risk analysis.
	
	\item \textbf{Blind Attack}: In this attack, the attacker does not know the number and type of devices installed in the smart home. We model blind attackers by using training data that is collected from a small scale IoT network containing a few popular IoT devices.  We then use  test sets which contain fewer or more types of devices than in the training set. For instance, a motion sensor might be in the test set but no in the training set.  The training set may contain all, some, or even none of the devices in the target smart home as the attacker does not know the type and number of devices installed in the target smart home. 
\end{itemize}

\subsection{Timestamp Obfuscation} \label{sub-sec:time-obfuscation}
In this section, we study the proposed timestamp obfuscation methods. Most of the  IoT devices generate transactions based on  a particular pattern which is reflected in the timestamp of the transactions. To make  device classification more difficult, we propose four multiple timestamp obfuscation methods which are as follows: 

\textit{1) Delayed Transactions:}  in this obfuscation method the transaction \textit{t} corresponding to a communication~\textit{c}  is generated with a random delay \textit{delay}. Thus, \textit{t.timestamp = c.timestamp + delay}  where  $ delay \in [0,d\textsubscript{max}] $. d\textsubscript{max} represents the maximum delay that is defined by the user. \textit{delay} value for each transaction is independent and is selected randomly. The aim is to obfuscate the pattern of transactions. This potentially may even change the order of transactions, i.e., $ t\textsubscript{i}.timestamp > t\textsubscript{i+1}.timestamp  $ where \textit{t\textsubscript{i}} is transaction corresponding to the \textit{i}th communication.

\textit{2) Multi-device Ledgers:} in this obfuscation method multiple devices share the same ledger to store their transactions. In conventional methods, each device employs the same ledger to  chain its transactions which potentially protects against Sybil attack where a malicious node pretends  to be multiple nodes. However, using the same ledger for each device  negatively impacts the user anonymity as changing the \textit{PK\textsubscript{+}} will not impact the anonymity as all transactions in a ledger belong to the same device which in turn facilitates  transaction classification. Thus, this obfuscation method stores transactions of multiple devices in the same ledger which in turn mixes the transaction temporal patterns. The transactions are stored in the ledger once a communication is made by the devices without any prioritization or limitation on the number of transactions each device can store.

\textit{3) Multi-packet Transactions:} in this obfuscation method each transaction may reflect multiple communications of a device. Depending on the device functions, the maximum time interval between two consecutive packets may vary, thus the exact number of combined packets depends on application and number of packets generated by the device. In this  obfuscation method the transaction can be considered as a summary of a number of communications made by the device which in turn reduces the volume of information available about communications of a device. By applying multi-packet obfuscation the pattern of transactions will not match with the device communication pattern which further complicates device classification.


\textit{4) Transaction Spoofing:} in this obfuscation method fake transactions, i.e., transactions that reflect no real communication, are added randomly to the ledger to mix the pattern of transactions and thus complicate device classification, e.g., storing a fake transaction after storing a transaction in the ledger. This method in turn increases the blockchain overhead as the size of the blockchain database increases.  The device randomly adds fake transactions in the blockchain to obfuscate the transaction  pattern.

To benchmark the performance of the outlined obfuscation methods, we consider a \textit{baseline} method where a single transaction is generated and stored in blockchain corresponding to a single communication. Having discussed the fundamentals, we analyze the implementation results next.  

\section{Evaluation and Results}\label{sec:evaluation}
In this section, we evaluate the proposed  timestamp obfuscation methods in Section  \ref{sub-sec:time-obfuscation}  through experiments on the empirical smart home dataset. We evaluate these methods for both informed and blind attacks. We first evaluate the obfuscation methods individually, then we study the impact of combining multiple obfuscation methods. 

\subsection{Delayed Transactions}
We consider three delay intervals which are  \([0, 0.5]\), \([0, 2]\) and \([0, 30]\) seconds and are selected based on the  separation time observed in  devices' patterns (see Table~\ref{tab:devices}).

\begin{figure*} [h]	
	\begin{center}
		\subfloat[]{\label{fig:informed-delay-day}
			\includegraphics[scale=.5]{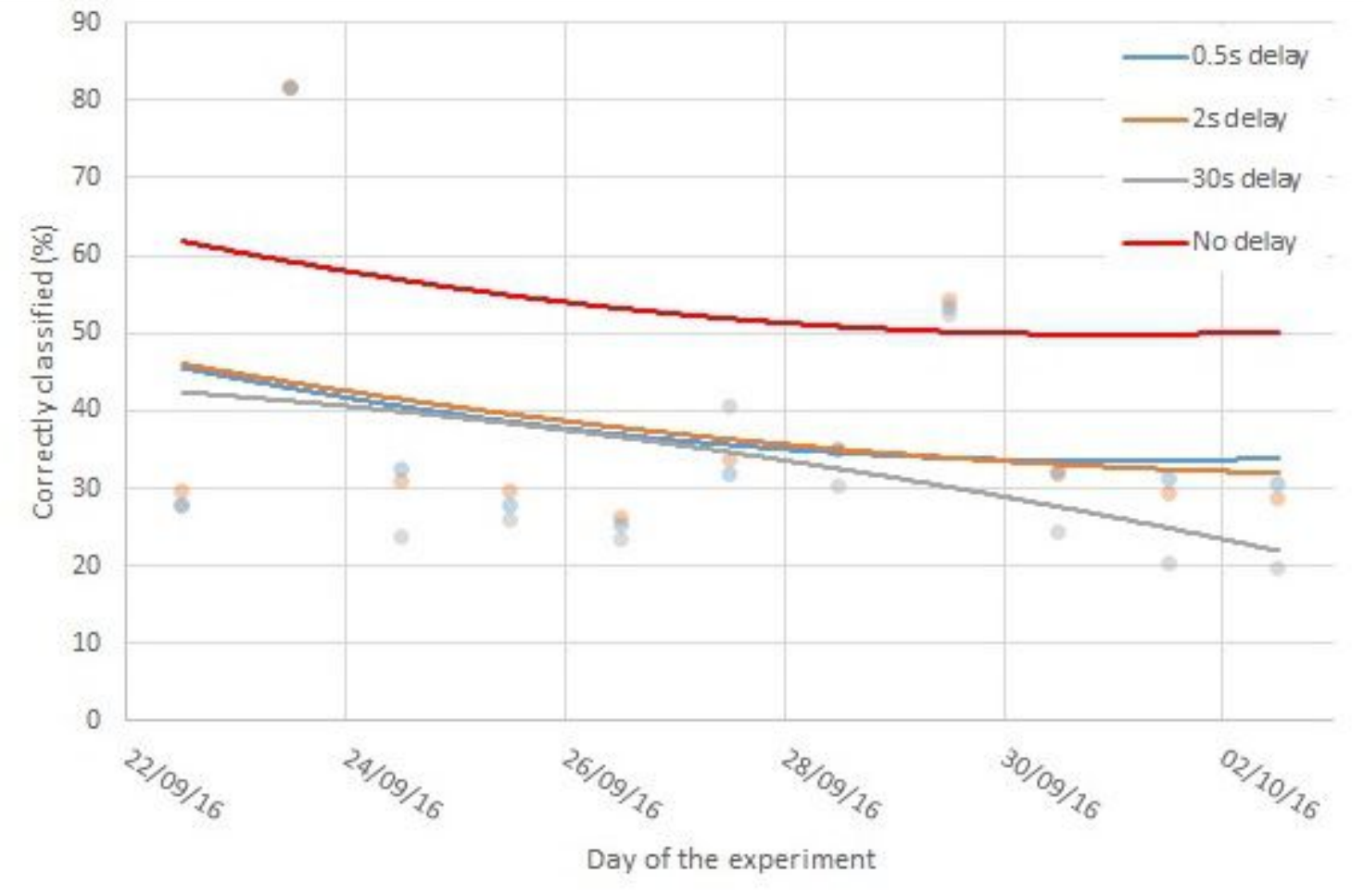} }
		\subfloat[]{\label{fig:blind-delay-day}
			\includegraphics[scale=.24]{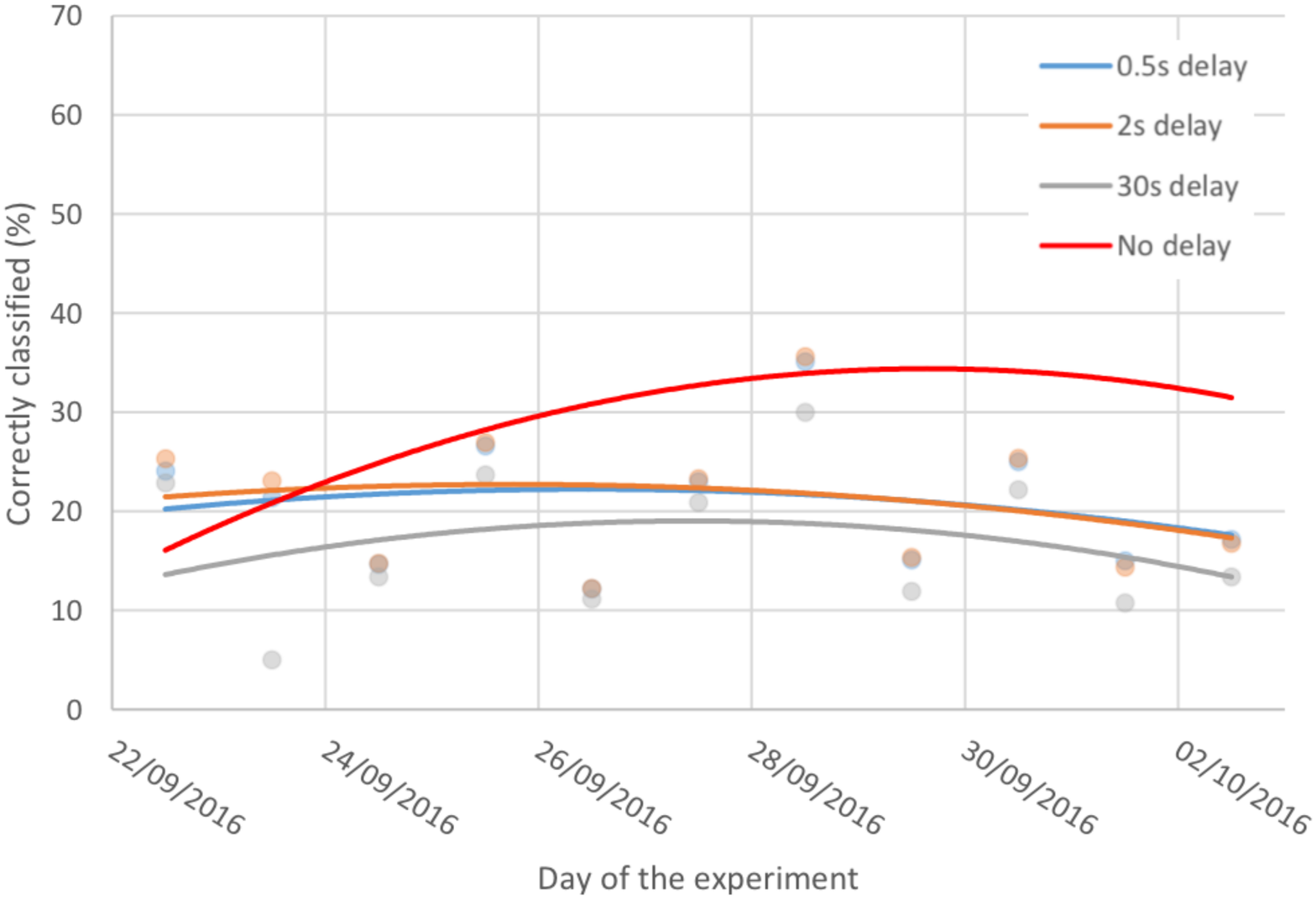} }
		\caption{The impact of delayed transactions for  informed attacks (a)  and blind attacks (b)(points represent error bar).}	\label{fig:delayed}
	\end{center}
	
\end{figure*}

\begin{figure} 	
	\begin{center}
	\includegraphics[scale=.5]{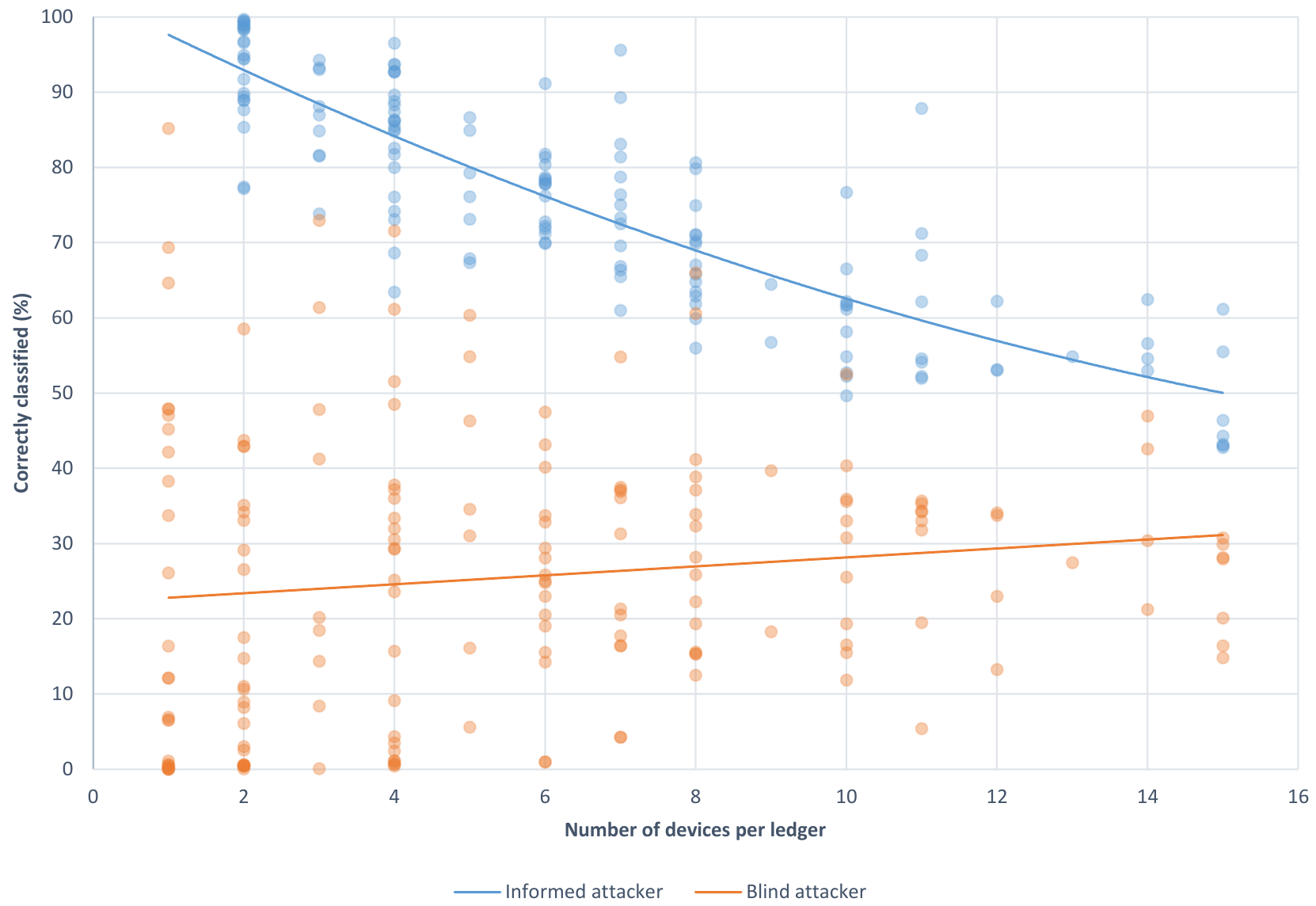} 
	\caption{The impact of multi-node ledgers in informed and blind attack models (points represent error bar).}
	\label{fig:multinode}
	\end{center}
	
\end{figure}

\begin{figure*} 	
	\begin{center}
		\subfloat[]{\label{fig:informed-delay-day} \includegraphics[width=9cm,keepaspectratio]{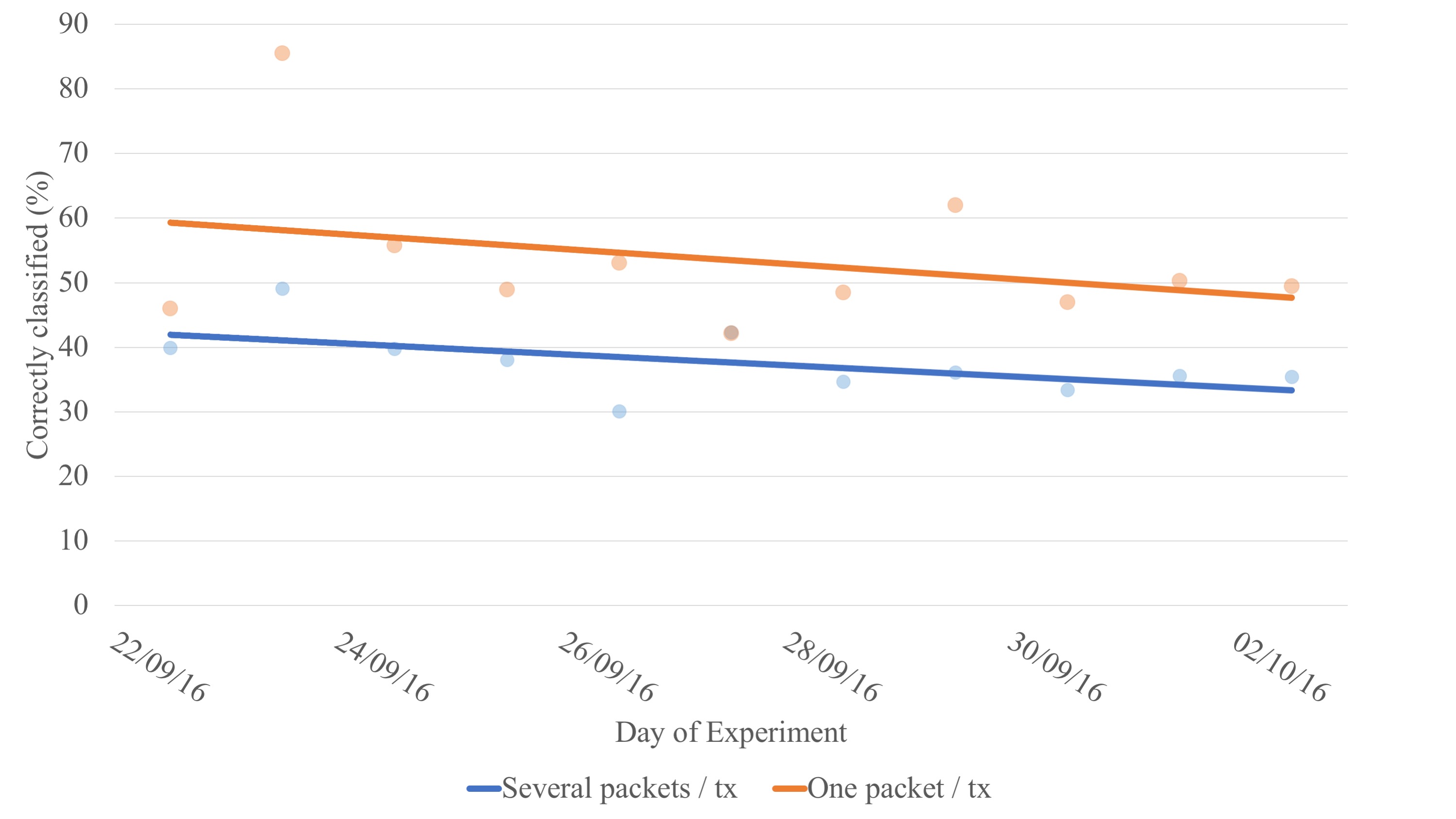} }
		\subfloat[]{\label{fig:blind-delay-day}
		 \includegraphics[width=9cm,keepaspectratio]{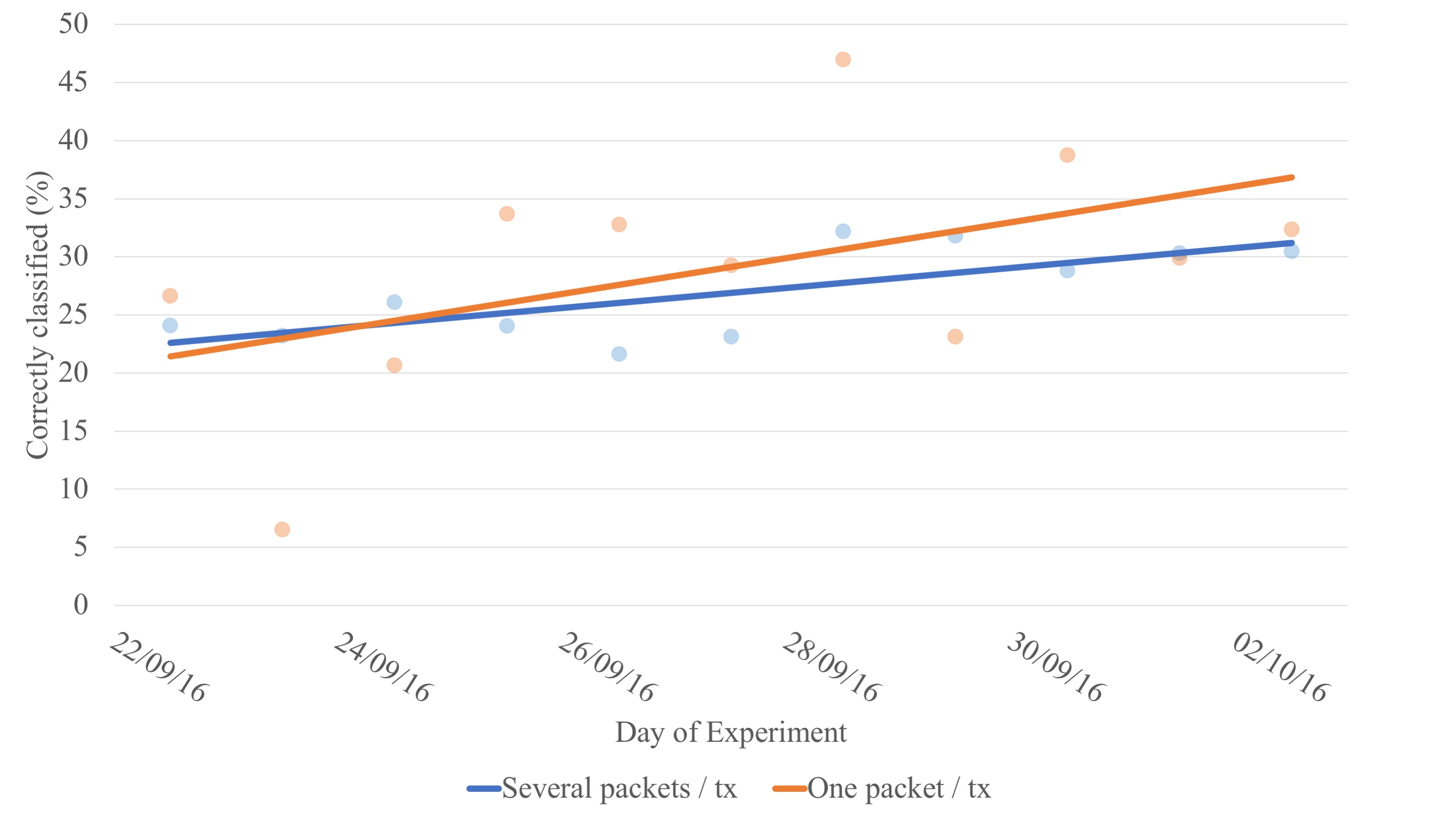} }
		\caption{The impact of multi-packet transactions for informed attacks (a) and blind attacks (b)(points represent error bar).}	\label{fig:multipacket}
	\end{center}
	
\end{figure*}

Figure~\ref{fig:delayed} shows the results for delayed transactions, for both informed  and blind attacks. In informed attacker model, the accuracy of classification is reduced by at least 15\% when applying the random delays. The results for \([0, 0.5]\) and \([0, 2]\) intervals have small differences given that the differentiable  times are either \(<0.4\)  or \(\geq 28\) seconds. However, changing the delay to 30 seconds makes it harder to classify transactions. For blind attacks, the classification accuracy drops for all approaches.  However, there is still a reduction of over 10\% with delayed transactions relative to the baseline approach. 

\subsection{Multi-device Ledgers}
As outlined earlier in Section \ref{sub-sec:time-obfuscation}, the devices are assigned randomly to common ledgers where the number of devices in each ledger varies. As evident from the experiment results shown  in Figure \ref{fig:multinode} the attack success rate reduced from 98\% for  ledgers used by single devices to around 50\% for ledger shared across 17 devices. With a small number of devices sharing the same ledger, the attacker can  distinguish transactions in the ledger by examining and matching transactions with pattern of known devices.  However, larger number of devices potentially reduces the success rate in linking transactions to a known device given the short time intervals between  transactions in the ledger. \par

The overall success rates for blind attacks is considerably lower, and interestingly, the general trend for average attack success rate slightly increases with more devices per ledger. In this attack model, we  train our algorithm to recognize an available subset of all possible devices, hoping that at least devices present in the home (the test) are in our training set. The test set can therefore  vary significantly from our training set. A large number of devices within the same ledger will make the test set more similar to the training set, so the classification accuracy increases on a large sample. When the number of devices per ledger are small, we observe a high variance in the results. This variance stems from the dependence of individual experiment runs on the specific selection of devices in the test set. Where there is a high match in test and training sets, the attacks can be quite successful and vice versa. We observe a trend of a drop in variance with an increase in the number of devices per ledger. Given this high variance across trials, we examine the maximum possible attack success in the model. Increasing the number of devices per ledger from 1 to 17 reduces the maximum attack success rate from around 85\% to 30\% for blind attacks. 
\begin{figure*} 	
	
	\begin{center}
		\subfloat[]{\label{fig:informed-delay-day} \includegraphics[width=9cm,keepaspectratio]{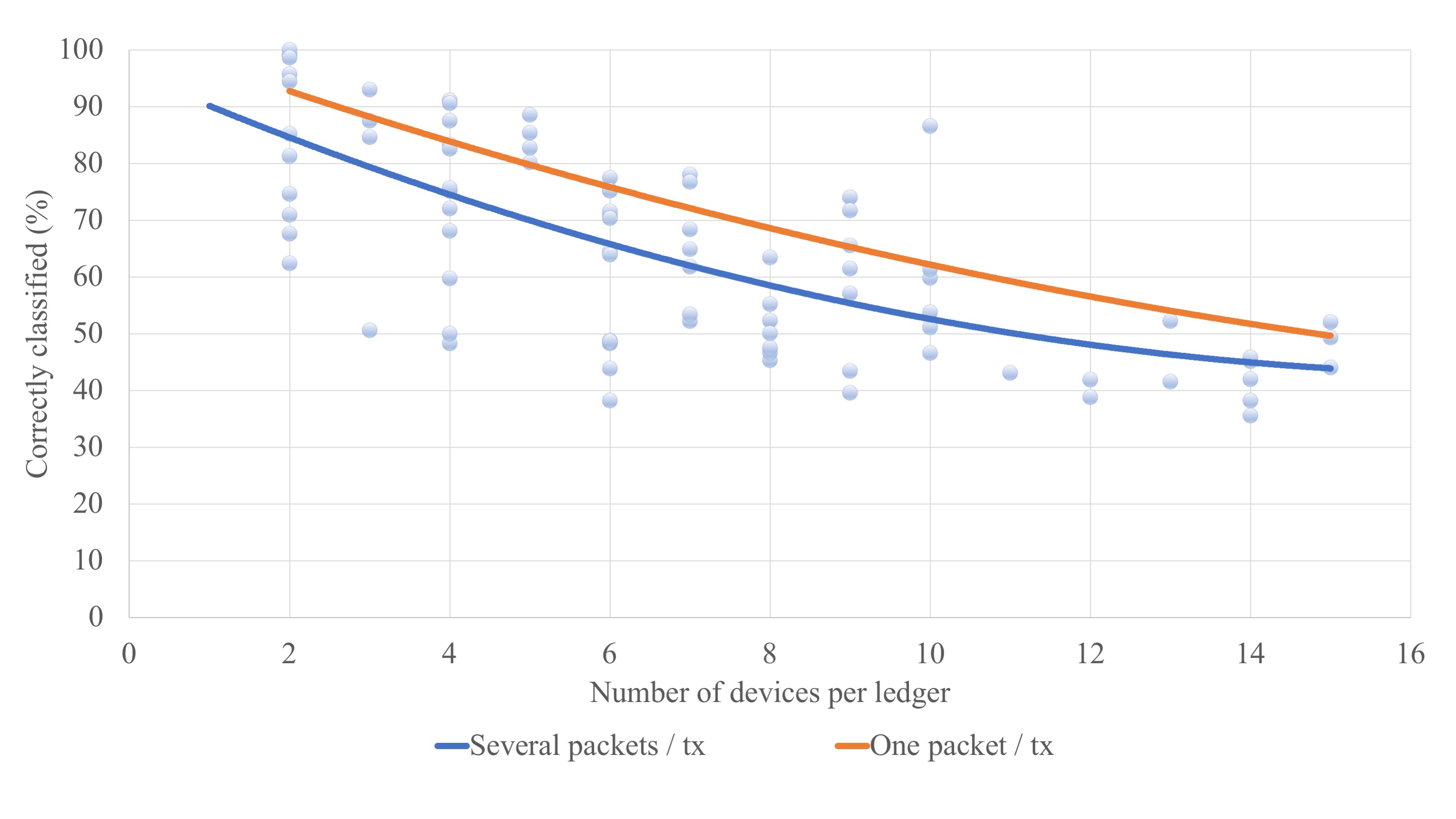} }
		\subfloat[]{\label{fig:blind-delay-day} \includegraphics[width=9cm,keepaspectratio]{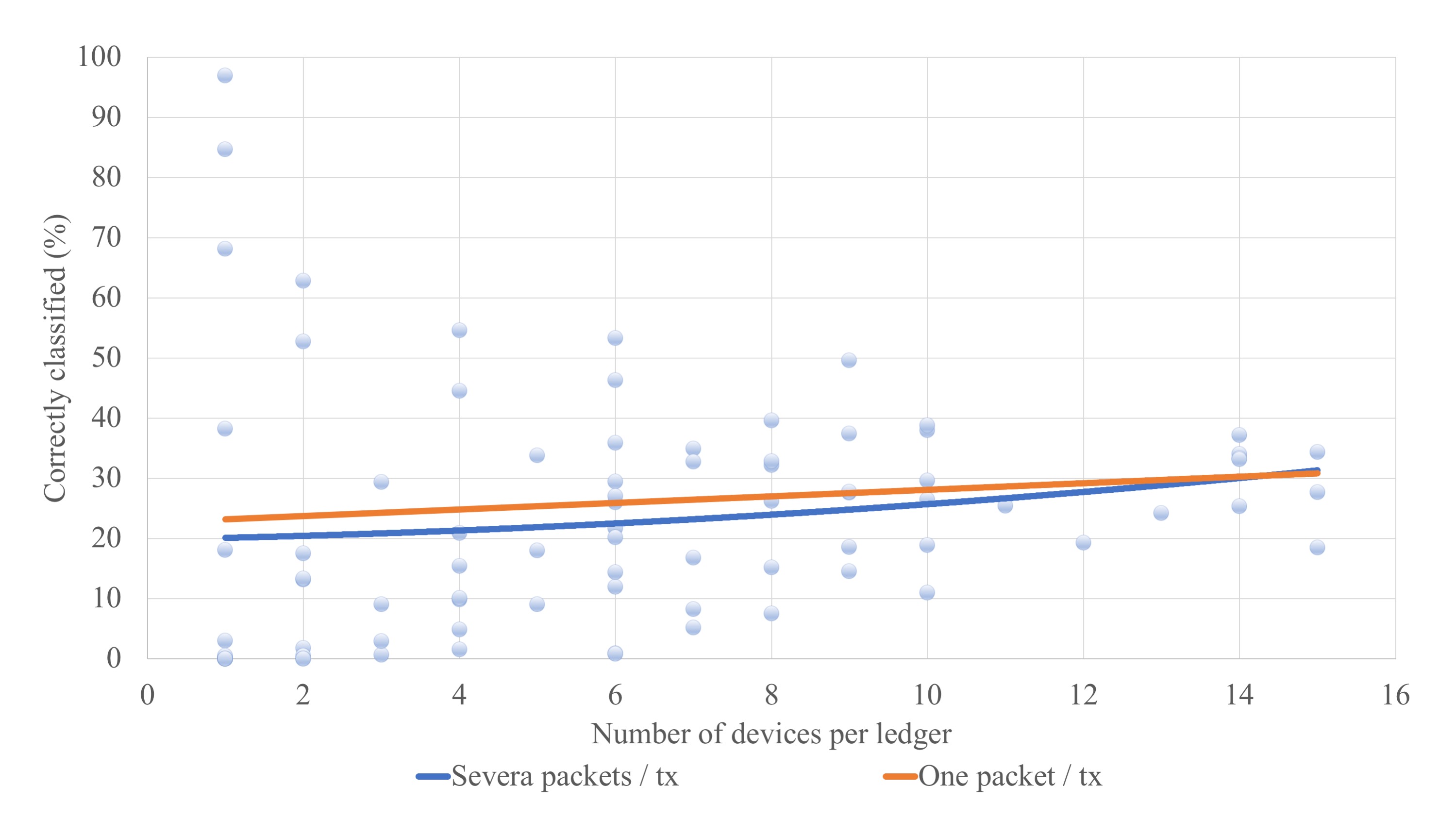} }
		\caption{The combined impact of multi-packet transactions and multi-node ledgers for informed attacks (a) and blind attacks (b)(points represent error bar).}\label{fig:multipacketledger}
		
	\end{center}
	
\end{figure*}
\begin{figure}
	\centerline{\includegraphics[width=9cm,height=9cm,keepaspectratio]{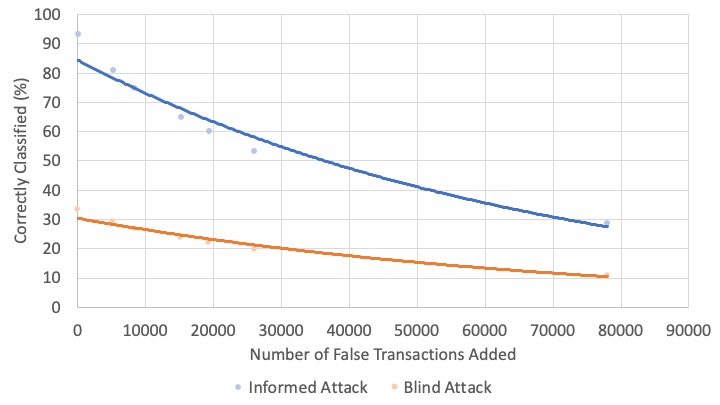}}
	\caption{The impact of transaction spoofing in blind and informed attacks (points represent error bar).}
	\label{fig:spoofing}
\end{figure}

\subsection{Multi-packet Transactions}
\begin{figure*} 	
	\begin{center}
		\subfloat[]{\label{fig:informed-delay-day} \includegraphics[width=9cm,keepaspectratio]{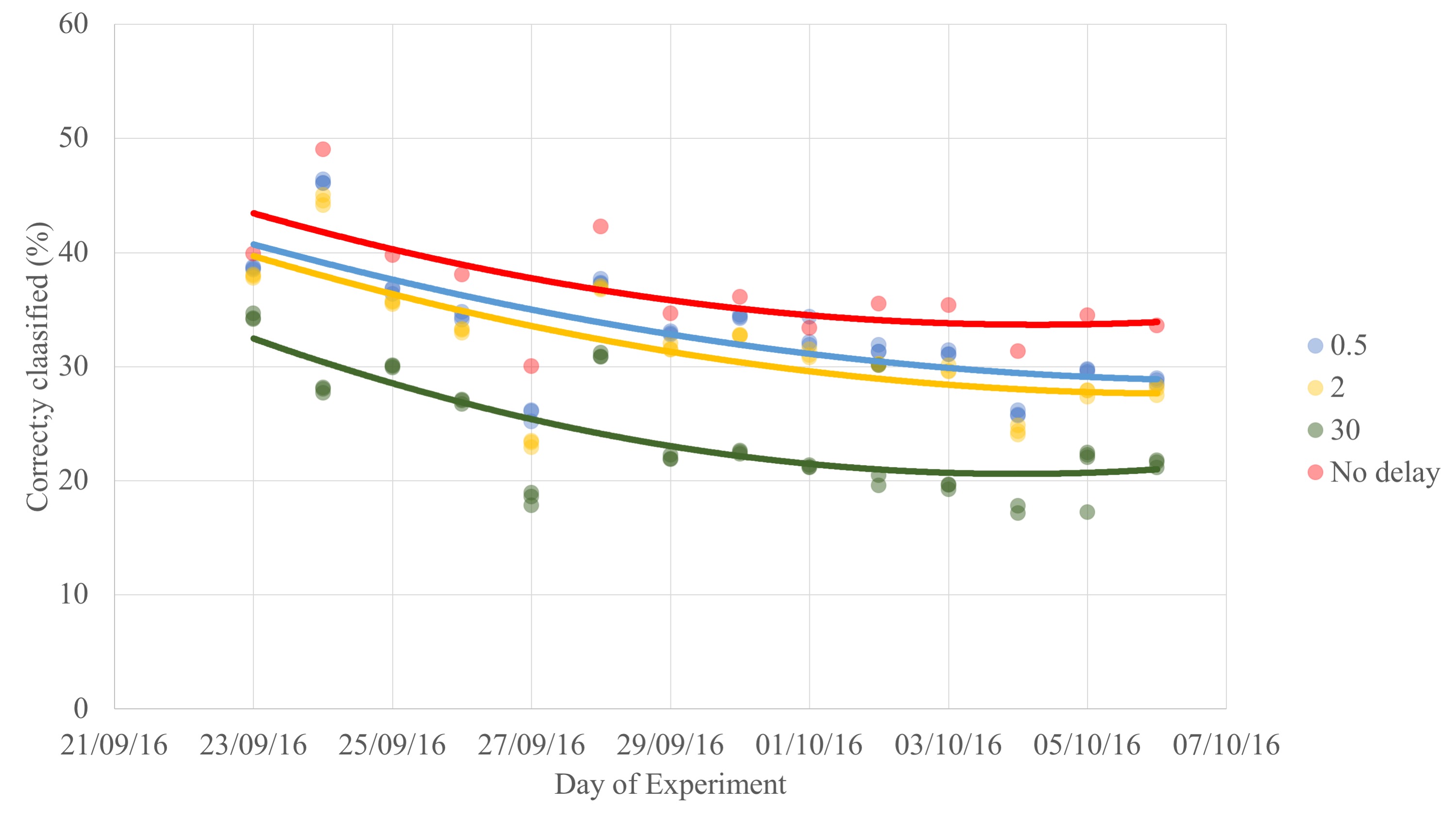} }
		\subfloat[]{\label{fig:blind-delay-day} \includegraphics[width=9cm,keepaspectratio]{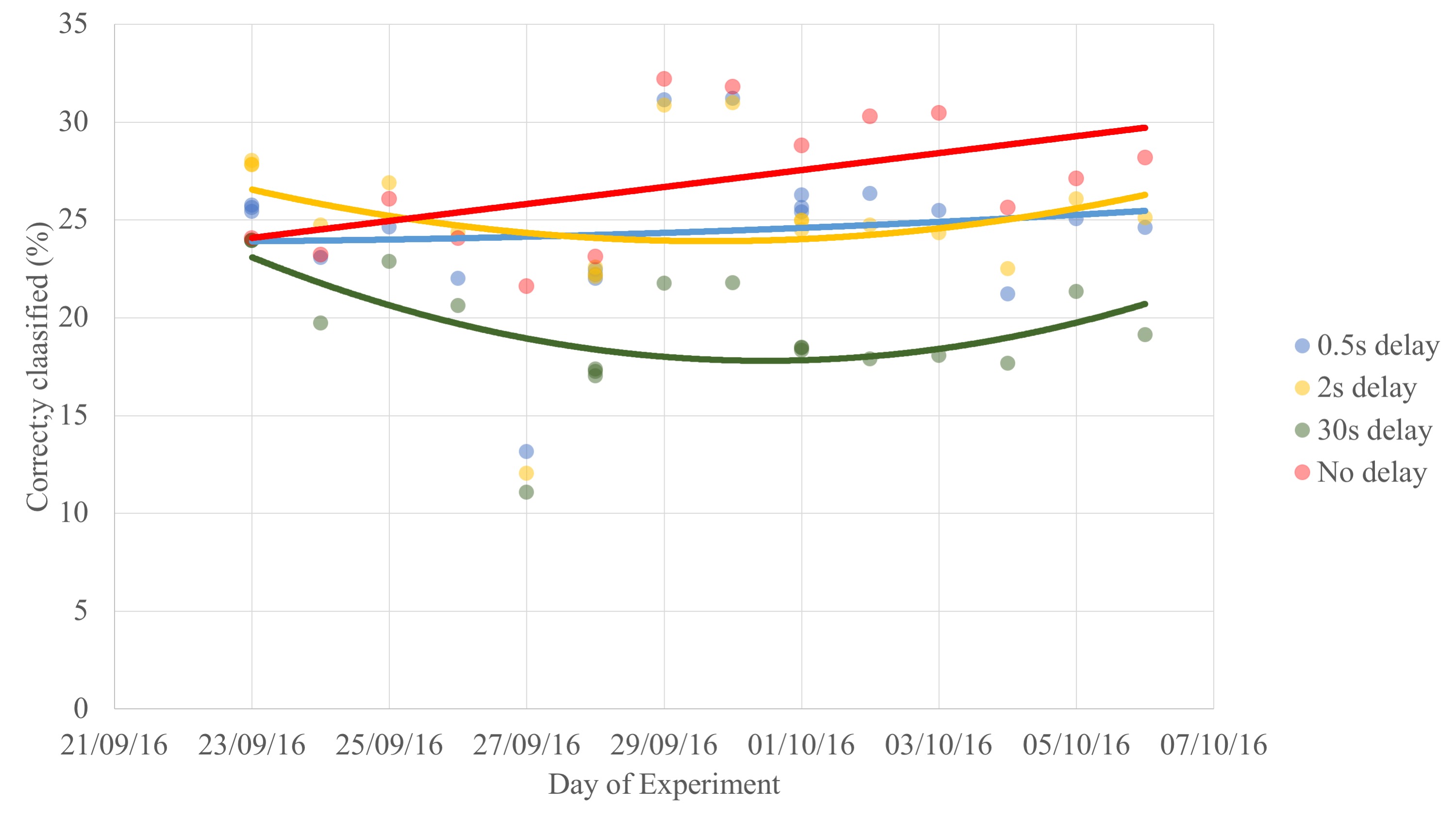} }
		\caption{The combined impact of multi-packet transactions and delayed for informed attacks (a) and blind attacks (b)(points represent error bar).}	\label{fig:multipacketledger-delayed}
	\end{center}
	
\end{figure*}
Next we explore the effects of merging multiple device communications into a single transaction. 
The experimental results for both informed and blind attacks are shown in Figure~\ref{fig:multipacket}. The classification accuracy decreases in the informed attack by 20\% on average. In one packet per transaction mode, the success rate in classifying transactions varies from 50\%-60\%.  Unlike one packet per transaction mode where the attacker  employs short separation times to distinguish devices, consolidating several packets in a transaction removes the separation time which in turn complicates device identification. On average, the success rate in classifying devices reduces by 20\% when employing multiple packet per transaction mode. \par 

The change is less obvious for blind attackers, but the accuracy observed with multi-packet transactions is lower than the baseline method. Recall that a blind attacker only uses a subset of actual devices in its training set, thus has lower visibility  to distinguish the devices, basing it only on higher separation times. Concealing the shortest separation times with multi-packet transactions has a greater impact on an informed attacker than on a blind attacker given the informed attacker's full visibility into the device type and their short separation times. There is significant variation in the differences between single and multi-packet transactions over different days in the simulation for blind attacks, as the performance depends significantly on the active devices on that day and whether these devices are in the blind attacker's training set.

\begin{figure*} 	
	\begin{center}
		\subfloat[]{\label{fig:informed-delay-day} 
		\includegraphics[width=9cm,keepaspectratio]{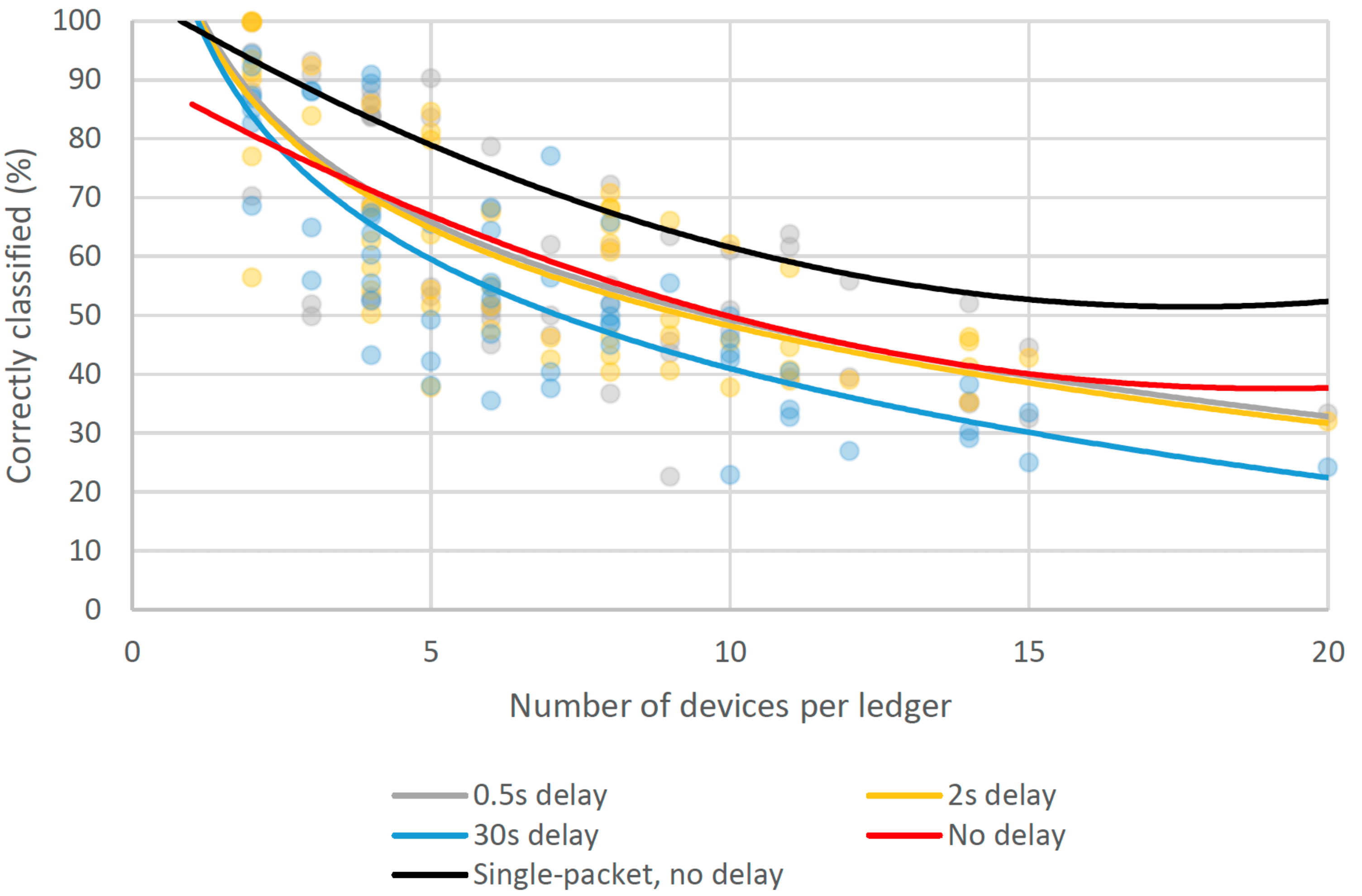} }
		\subfloat[]{\label{fig:blind-delay-day} 
		\includegraphics[width=8.5cm,keepaspectratio]{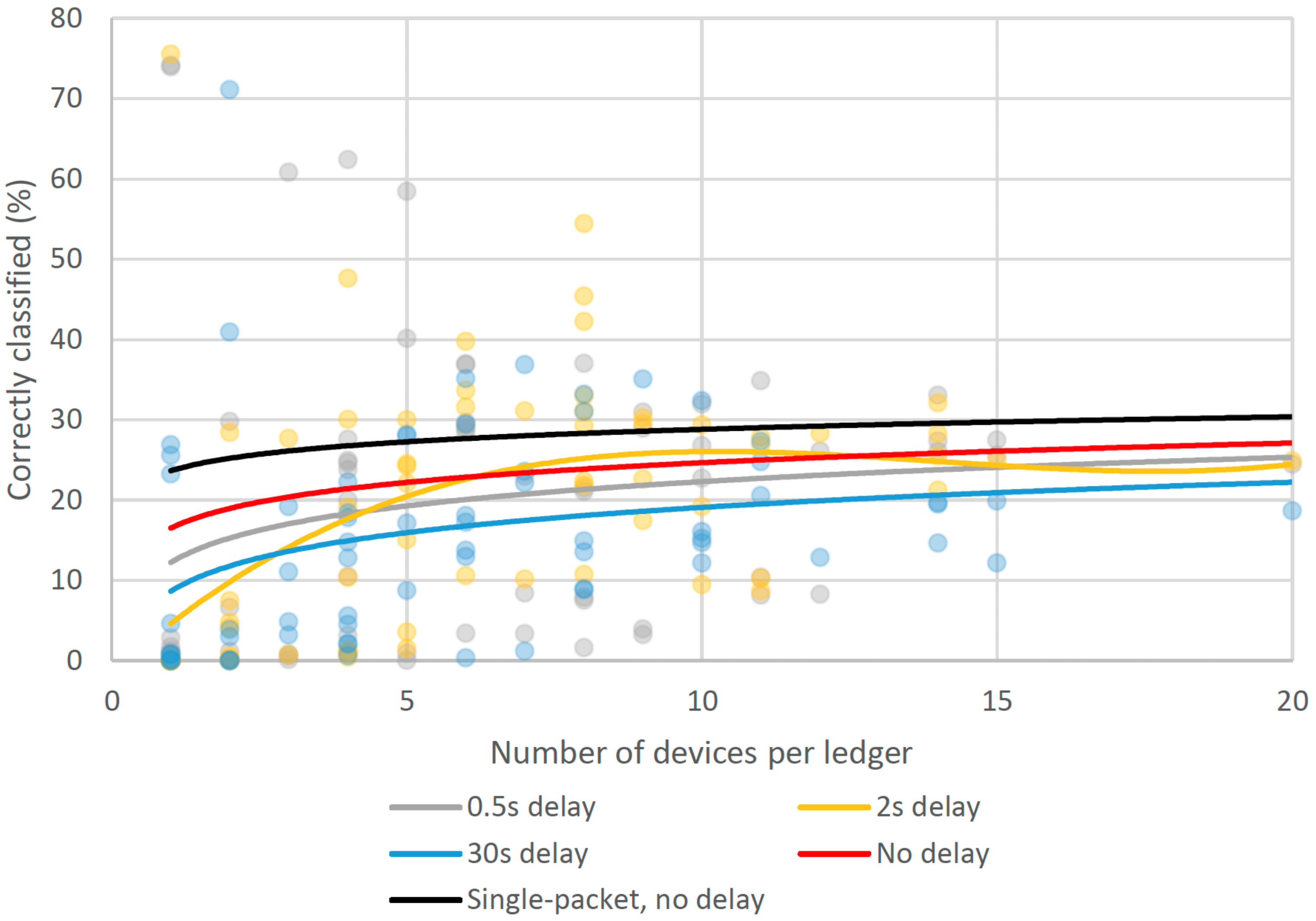} }
		\caption{The combined impact of combining three methods for informed attacks (a) and blind attacks (b)(points represent error bar).}	\label{fig:combined}
	\end{center}
	
\end{figure*}


\subsection{Transaction Spoofing}
Finally, we quantify the effect of adding fake transactions to obfuscate the temporal pattern of genuine transactions. 
The experimental results for both informed and blind attacks are shown in  Figure \ref{fig:spoofing} that shows the decrease in the classification accuracy in both blind and informed attacks based on the different number of spoofed transactions added to the ledger. A higher number of spoofed transactions results in reduced classification accuracy for an attacker as it is harder to recognize the timestamp pattern. For the informed attack, the success rate is reduced from 93\% to 28\%, and the blind attack success rate is reduced from 33\% to 10\% for 3 times the original amount of transactions added. With more false transactions in between the original timestamps, the temporal patterns are mixed up making device identification, based on pattern matching, more challenging. The improved anonymity comes with the cost of more transactions to be committed to the blockchain and thus more overhead in the blockchain. \par

\subsection{Combined Timestamp Obfuscation}
In the previous subsections, we studied the impact of applying  each timestamp obfuscation method separately. As the number of devices in the dataset is about 20 devices, the percentage of correctly classified transactions considering a strategy that selects devices randomly should remain around 5\%. In this subsection, we evaluate the impact of combining the proposed obfuscation methods.

First, we study the impact of combining multi-packet transactions with multi-node ledgers. As shown in Figure \ref{fig:multipacketledger}, in the informed attack the success rate of classifying the devices is reduced by about 10\%. Combining multi-packet transactions with multi-node ledgers yields results extremely close (less than 5\% of difference)\, to the ones obtained with single-packet transactions, for blind attackers. Again, the reason is that blind attackers rely more on longer packet separation time and are less impacted by obfuscations affecting short separation times.  

We next evaluate the impact of combining  delayed transactions with multi-packet transactions (cf. Figure~\ref{fig:multipacketledger-delayed}).  
The three delay intervals have the same impact as studied in Figure~\ref{fig:delayed} by 30 sec interval having the most visible improvement. Comparing the results in Figure~\ref{fig:multipacketledger-delayed} with Figure~\ref{fig:delayed}, we noticed the success rate of device classification is lower in delayed single-packet per transaction mode. This can be due to the balanced distribution of transactions in the dataset. The transaction distribution significantly changes by applying  multi-packet per transaction mode. This can cause  some devices to  dominate the ledger leading to a higher success rate in identifying the devices (as the attacker identifies more dominant devices).

Now we evaluate the impact of combining delayed transactions, multi-packet transactions, and multi-node ledgers (cf. Figure~\ref{fig:combined}). 
In an informed attack, the best performance is achieved by the combination of all three methods with a delay of 30 sec that reduces the classification accuracy to 24\%.  For blind attacks, there is a slight advantage for the combined approach with 2 second delays for ledgers of 1-2 nodes, while again overall the combined approach with a 30 second delay achieves the attack success rate at 19\%.

\section{Conclusion and Future Work}\label{sec:conclusion}
In this paper, we examined the significant issue of IoT device identification. We have quantified the impact of attacks in device de-anonymization. That is the re-identification of a device/pattern based on the previous interactions.  We used two attack models, namely informed and blind, to test the attacks scenarios against four times-stamp obfuscation methods. For that purpose, we used a data-set of a smart home network traffic consisting of 20 IoT devices over two weeks. For the evaluation, we first used the obfuscation methods individually, then investigated the impact of multiple obfuscation methods together to see the differences in adding complexity. Our experimental results showed that the success rate of the device classification reduced significantly with the employment of the combination of the times-stamp obfuscation methods for both the informed and blind attacks.\par
As future research direction, additional obfuscation methods can be introduced to reduce the classification rate while considering the overheads. The obfuscation methods introduced in this paper reduce the sucess rate of classifying attacks. However, they potentially modify the pattern of transactions in blockchain that may potentially impact services a user receives based on the blockchain transactions. Studies are required to allow users to re-create ordered list of their transactions.
This paper studied smart home as an application. The impact of user/device deanonymization can be studied in other application domains including smart grids.

\balance{}

\bibliographystyle{IEEEtran}
\bibliography{bare_jrnl_compsoc}
\end{document}